\documentclass[showpacs,twocolumn,amsmath,superscriptaddress,prl]{revtex4}

\usepackage{hyperref} 
\usepackage[dvips]{graphicx}
\usepackage{epsfig}
\usepackage{color}
\usepackage{subfigure}
\usepackage{soul}
\usepackage{float}

\begin{document} 
\title{
Heaps' Law and Vocabulary Richness\\
in the History of Classical Music Harmony\\
}
\author{Marc Serra-Peralta}
\affiliation{%
Centre de Recerca Matem\`atica,
Edifici C, Campus Bellaterra,
E-08193 Barcelona, Spain
}
\author{Joan Serr\`a}
\affiliation{Dolby Laboratories, Diagonal 177, P10,
E-08018 Barcelona, Spain}
\author{\'Alvaro Corral$^\dag$}
\affiliation{%
Centre de Recerca Matem\`atica,
Edifici C, Campus Bellaterra,
E-08193 Barcelona, Spain
}\affiliation{Departament de Matem\`atiques,
Facultat de Ci\`encies,
Universitat Aut\`onoma de Barcelona,
E-08193 Barcelona, Spain
}\affiliation{Barcelona Graduate School of Mathematics, 
Edifici C, Campus Bellaterra,
E-08193 Barcelona, Spain
}\affiliation{Complexity Science Hub Vienna,
Josefst\"adter Stra$\beta$e 39,
1080 Vienna,
Austria
}
\email{alvaro.corral@uab.es}

\begin{abstract} 
Music is a fundamental human construct, 
and harmony provides the building blocks
of musical language. 
Using the {\it Kunstderfuge} corpus of classical music, 
we analyze the historical
evolution of the richness of harmonic vocabulary of 76 classical composers, 
covering almost 6~centuries. 
Such corpus comprises about 9500 pieces, 
resulting in more than 5 million tokens of music codewords. 
The fulfilment of Heaps' law for the relation between the size of the harmonic vocabulary of a composer (in codeword types) and the total length of his works (in codeword tokens), with an exponent around 0.35, allows us to define a
relative measure of vocabulary richness that has a transparent interpretation. 
%
%
When coupled with the considered corpus, this measure allows us to quantify harmony richness across centuries, unveiling a clear increasing linear trend. 
In this way, we are able to rank the composers in terms of richness of
vocabulary, in the same way as for other related metrics, such as entropy.
We find that the latter is particularly highly correlated with our measure of richness.
Our approach is not specific for music 
and can be applied to other systems built by tokens of different types, 
as for instance natural language.
%
%
%
\end{abstract} 

\date{\today}

\maketitle

\section{Introduction}



It is a well-known but nevertheless 
intriguing fact that the usage of natural language,
as reflected in texts or speech,
displays very strong statistical regularities \cite{Altmann_Gerlach,Hdez_FCancho_book,Torre19,Corral_brevity}.
Although the most popular and in-depth studied of these
is Zipf's law of word frequencies
\cite{Baayen,Baroni2009,Zanette_book,Piantadosi,Moreno_Sanchez}, 
the most fundamental linguistic statistical law is probably 
Heaps' law, also called Herdan's law \cite{Baeza_Yates_Ribeiro,Kornai2002,Font-Clos2013,Corral_Font_Clos_PRE17}. 
This law relates the two main quantities that are necessary
to set the statistical analysis of a text:
the number of word tokens in the text, i.e., 
its length (in words), $L$,
and
the 
number of word types, 
which is referred to as the size of the vocabulary of the text, $V$. 
More precisely, 
Heaps' law 
states that the relation between $V$ and $L$ is reasonably well approximated by a power law,
$$V \simeq K L^\alpha,$$ 
with an exponent $\alpha \in (0,1]$
and a proportionality constant $K$.

Care has to be taken though when referring to Heaps' law, 
as there are in fact two versions of it and considerable confusion 
between
them.
The first version 
deals, in principle, with just one
text, 
and studies the growth of the accumulated vocabulary size as the text 
is read, 
from beginning to end.
This yields a non-decreasing curve with no scattering that 
ends in the total values of 
vocabulary size and length
for the considered text
\cite{Mandelbrot61,Heaps_1978,Serrano}.
Although there are some derivations that relate this version of Heaps' law with Zipf's law,
the situation is not so simple, 
and this version of Heaps' law is usually a bad description of type-token growth,
even if some formulation of Zipf's law can be considered to hold
\cite{Font-Clos2013,Font_Clos_Corral}.
Instead of a linear trend in a log-log plot, what is usually observed
is a slightly convex shape
that has sometimes been confounded with a saturation effect \cite{Font_Clos_Corral}.
%
%

The second version of Heaps' law, 
the one considered in this study,
needs a number of texts, or a collection of documents, 
and compares the total (final) values $V$ and $L$
for each one of the complete texts or documents. 
This results in a scatter plot, to which a power-law curve can be fitted to account for the 
correlation between $V$ and $L$.
This version of the law can be justified using the generalized central limit theorem 
\cite{Corral_Font_Clos_PRE17}, 
which has the advantage that does not require that Zipf's law holds exactly,
but only asymptotically. 
Overall, one must distinguish between an intra-text Heaps' law
(type-token growth) and an  
inter-text Heaps' law (vocabulary-length correlation).



In practical applications, one is usually interested in quantifying 
attributes of the entities under study, 
and such quantification should be grounded on well-established statistical laws. 
In the case of natural language,
there is an old tradition of evaluating richness of vocabulary~\cite{Wimmer_Altmann}, 
which 
has direct applications in authorship and genre analysis \cite{Kubat}. 
%
%
%
Comparing different texts (or authors),
one could wrongly associate size of vocabulary $V$ with richness of vocabulary.
But the longer the text, the larger the vocabulary (on average),
and texts of different length $L$ cannot be compared in this way.
One simple solution \cite{TTR} is to divide the vocabulary size by the text length, 
yielding the so-called type-token ratio, $V/L$.

Nonetheless, 
the fulfilment of 
Heaps' law tells us that this linear rescaling is unjustified
and, at the same time, suggests a natural solution.
Several indices have been proposed following this reasoning
(also as variations or alternatives to Heaps' law \cite{Wimmer_Altmann});
for instance, Guiraud's index, 
$$I_G=\frac V {\sqrt{L}},$$
or Herdan's index,
$$
I_H=\frac{\log V}{\log L}.
$$
The former assumes Heaps' law with the ``universal'' value $\alpha=0.5$ (and $I_G=K$),
while the latter fixes $K=1$ (with $I_H=\alpha$).
The choice of $K=1$ is naively justified by the obvious fact that $V=1$ when $L=1$;
nevertheless, one should interpret Heaps' law as a large-scale emergent property of texts that does not necessarily describe
small-scale behavior. 
Note also that both indices are ``absolute'', 
in the sense that they can be obtained for a single text or document
(as one of the parameters of Heaps's law, either $K$ or $\alpha$, 
are considered universal and fixed a priori).



As language, music is an attribute that defines us as humans 
\cite{Ball_music}.
A number of authors have pointed out similarities between language and music, and some have argued that music is indeed a ``language'' \cite{Zanette_music}, 
although the notion of grammar and semantic content
in music has been debated \cite{Zanette_music,Ball_music}. 
%
Despite this, it is clear that 
there exist strong relations between language and music, specially regarding rhythm, pitch, syntax, and meaning \cite{Patel}.
This way, one can consider music as a succession (in time) of some musical symbols 
(which can be considered analogous to words in texts or speech), 
for which statistical analysis can be performed in the same way 
as in quantitative linguistics
\cite{Zanette_nature08}.
%

%
However,
a remarkable problem is that, in contrast to language, 
the individual entities to analyze in music are not immediately clear
 \cite{Corral_Boleda}.
For instance,
Manaris et al. \cite{Manaris1} mention diverse possibilities
involving different combinations of pitch and duration,
as well as pitch differences.
This, together with some difficulties to deal with musical datasets in an automatized way,
may explain the fact that the study of 
linguistic-like laws in music has been 
rather limited,
in comparison with the study of natural language.
Nevertheless, let us point out to 
Refs. \cite{Zanette_music,Beltran,Haro,Serra_scirep,Liu}
as some of the pioneering analyses exploring
the applicability of Zipf's law in music
(providing weak evidence in some cases).

In this paper, we deal with the suitability of statistical laws,
in particular Heaps' law,
to describe regularities and patterns in musical pieces
and 
to provide a natural metric to assess harmonic vocabulary richness.
In contrast to previous metrics, this new one is not absolute, but relative 
to a given corpus or collection, as
we argue that richness is more properly defined relative to an underlying probability distribution.
Moreover, such relativeness
yields a considerable advantage of interpretability of the obtained values. 
We also compare this metric with more common measures,
finding that it is correlated with 
the entropy of the type-count distribution.
To perform this research,
we analyze classical music as captured by MIDI musical scores
\cite{Wikipedia_MIDI}
and construct music codewords in a way similar to the one of Ref. \cite{Serra_scirep}.
%
%

In the next sections,  
we describe the extraction of music codewords from MIDI files (Sec.~2),
overview the music corpus used (Sec. 2 too),
present empirical results for Heaps' law (Sec. 3),
develop and apply our metrics for the quantification 
of vocabulary richness (Sec. 4),
compare with entropy and codeword filling (Sec. 5),
and test the robustness of the results
in front of the criteria employed in codeword construction (Sec. 6).
Naturally, we end with some discussion and conclusions (Sec. 7).
Our work can be put in the context of {\it culturomics},
the rigorous quantitative
inquiry of social sciences and the humanities
by means of the analysis of big-data \cite{culturomics}.
To facilitate assessment and future work on the topic, 
the code used in this paper is available at 
\url{https://github.com/MarcSerraPeralta/chromagramer}.

\section{Processing and data} 

\subsection{Construction of music codewords}

In contrast to the famous composer Olivier Messiaen 
(and others), 
who considered rhythm as the essence of music, 
we put our focus on 
harmony,
understood here simply as the combination of pitches 
(or, more precisely, pitch classes, as we will see immediately)
for a given time frame.
Our starting point are MIDI files, 
with each MIDI file corresponding to 
one electronic score of a musical piece
(the corpus used is described at the end of this section).
MIDI stands for Musical Instrument Digital Interface
(for more information see the bibliography in Refs. 
\cite{Wikipedia_MIDI,Benson_music_book}).

One might complain that audio recordings 
are richer than MIDI files as expressions
of music, as the latter may lack the variability and nuances of interpretation
\cite{Geisel_music}
(some other MIDI files may additionally not correspond to ``original'' scores, 
but are created from the life performance of a musical piece).
However, we may argue that a musical score contains 
the core of a piece.
As we will need to discretize the elements that form the pieces,
scores provide an objective first step in such discretization.
The situation may be considered similar to the relation between written language and speech \cite{Torre_Lacasa,Torre19}.
Moreover, as no original recordings exist for classical music
(except for the last 100 years or so), 
scores are our 
best
link with the spirit of the composer.

\begin{figure}[ht]
\includegraphics[width=.85\columnwidth]{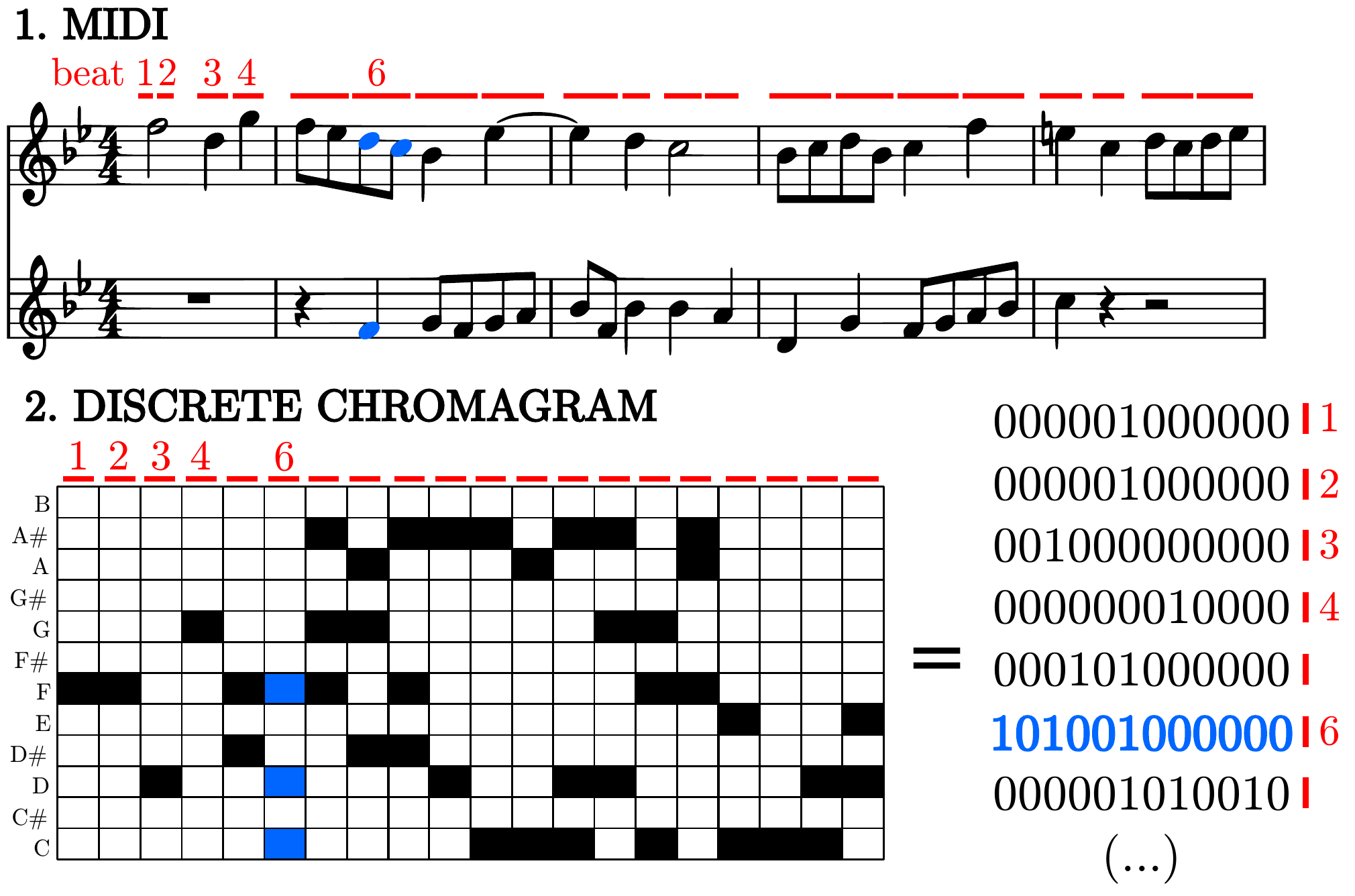}
\caption{
Example of a sequence of discretized chromas (chromagram) arising
from a MIDI score. 
Note that the two pentagrams constituting the score have to be read in parallel.
}
\label{Figexample}
\end{figure}

The procedure to obtain elementary units
(codewords) from MIDI files is the following 
(similar to the one in Ref. \cite{Serra_scirep},
and illustrated in Fig. \ref{Figexample}):

\begin{itemize}
\item
First, 
MIDI files are read using the program {\tt midi2abc}
\cite{midi2abc}
and converted into standard text files
containing the onset time, duration, and pitch of each note.
Pitches range from
$C$-$1$ (fundamental frequency at 8.1758 Hz), MIDI note 0,
to $G9$ (12,543.85 Hz), MIDI note 127.
$A4$ (440 Hz) is MIDI note 69. 
For some pieces starting and/or ending with silences (empty bars or beats), 
these are removed, in order to avoid an 
artificial overestimation of $L$. 
The files also contain metadata about the key of the pieces, 
which we disregard (due to some unreliability we found in 
some control checks in the corpus we used).


\item
With the purpose of reducing dimensionality, 
all notes are collapsed into a single octave, i.e.,  
the pitches $A0$, $A1$, $A2$,
etc., 
are considered to be the same note, or pitch class, $A$ 
(or {\it la}, in solf\`ege notation), 
and the same for the rest of pitches.
This leads to 12 pitch classes.
This dimensional reduction has also a perceptual basis,
rooted on the prevalence of the Western reference system based on the octave
\cite{Ball_music,Krumhansl1982}.
 
\item
In order to decrease temporal resolution,
each score is divided into small discrete time frames.
By default, 
we chose this time unit to be the beat
(as this is probably the most relevant time scale in music \cite{Ball_music}).
For example, 
the $^4_4$ bar yields $4$ beats per bar,
whereas  
the $^3_8$ bar yields $3$.




\item
For each time frame  
we construct a 12-dimensional vector,
with one component for each pitch class, 
$C,C\#,\dots G\#, A,A\#,B$,
i.e., from {\it do} to {\it si}.
Each component of the vector contains the sum 
of all durations of notes of the corresponding pitch class in the corresponding time frame. 
All notes coming from different instruments (or different pentagrams) playing in parallel are counted at a given time frame (as seen in Fig. \ref{Figexample}),
due to the fact that they are perceived together by the listener.
In this way, if the unit is the beat, 
in the $^4_4$ bar, 
each quarter note or crotchet counts as one, 
and each eighth note or quaver counts as one half, 
whereas in the $^3_8$ bar, 
it is the eighth note which counts as one.
Notes that occupy more than one time frame
are split according to their duration in each of them
(and thus, the maximum contribution coming from an individual note in the score 
is one at each time frame).
The vectors obtained in this way are called chromas.


\item
To further simplify, and in order to get well defined countable entities, 
chroma vectors are discretized,
with components below a fixed threshold (0.1 by default) reset to zero
and components above the threshold reassigned to one.
In this way, a value equal to one in a component of the discretized vector means that
the corresponding pitch class has a significant presence (above 0.1) in that time frame,
whereas a value of zero means that the pitch class has null or very little weight
and can be disregarded.
We refer to the resulting discretized vectors as discretized chromas.


\end{itemize}

Note that the procedure of codeword construction has just two parameters:
the time unit, which we take to be the beat, 
and the discretization threshold, which we have equated to 0.1.
These values can be considered somewhat arbitrary (in particular the threshold).
Therefore, to demonstrate the generality of our approach, 
we also test the robustness of our results in front of different prescriptions for
them (Sec. 6).


Discretized chromas constitute our music codewords, and 
turn out to be nothing else than 12-digit binary numbers 
(from 0 to $2^{12}-1=4095$, which also represent the
vertices of a 12-dimensional hyper-cube).
One could argue that these entities would be closer to music letters
than to music words; however, previous statistical analysis \cite{Serra_scirep}
shows that the complexity contained in them is enough to treat them as music words.
Further, our codewords have ``temporal structure'', 
as they can be composed by the succession of several shorter notes
(for example, in the $^3_4$ bar, the duration of one beat is that of two consecutive quavers, or four semiquavers, etc.).
Perhaps, the fundamental difference with words
is that our music codewords have a fixed length 
(that of the selected time unit), 
whereas in linguistics the length of words is variable 
(given by the number of letters, for instance)
and determinant for the validity of Zipf's law \cite{Corral_brevity}. 

\subsection{Transposition}
An 
additional
step of the procedure is to transpose all pieces to a common key,
so that all major keys are transposed to $C$ major
and all minor ones to $A$ minor
(the reason of using $A$ minor instead of $C$ minor is that 
the former is the relative minor of $C$ major, 
sharing the same key signature and leading thus to a more similar usage of chromas).
For example, if
a piece 
is in $G$ major, 
all $G$ pitches in the piece are transformed to $C$ pitch class, 
all $G\#$ pitches to $C\#$, and so on.
This is just a shift in pitch-class space.
Although keys 
are directly provided by the MIDI file
(at least for the majority of files),
an initial inspection showed that this information is unreliable,
and thus we perform our own analysis of key
 (see Appendix II for details).
No attempt is performed to identify changes of key inside a piece,
so we calculate the predominant or more common key for each piece. 


If a piece keeps a constant key, 
transposition has no influence 
on the size of its vocabulary,
but if different pieces are merged into a single dataset (as we will do),
it can be convenient to merge them after they are transposed to the same key.
In this case, if the pieces come in different keys, 
transposition leads obviously to a different vocabulary,
where one does not deal with pitch classes but with tonal function
(and thus, the resulting $C$ pitch class represents the tonic, 
$G$ represents the dominant, etc. \cite{Grove,Music_dummies}).
Transposition can be useful also to unveil a reduced vocabulary, 
where a given composer could show what seems a broad apparent richness that
arises from a limited vocabulary transposed into a number of different keys.
%


In some sense, for listeners with absolute pitch \cite{Ball_music}, 
it makes sense to keep the original key of different pieces 
when they are merged to form a larger dataset 
(these listeners naturally distinguish the different pitches and also the keys).
However, for the vast majority of listeners (those with relative pitch), 
it is more natural to merge pieces 
while transposing 
to a fixed key
(at least in the usual equal-temperament tuning system \cite{Ball_music}).
For this reason, we will 
only work with transposed pieces
in our analysis.

\subsection{Concatenation and elementary statistics}

As we are particularly interested in studying individual composers,
we concatenate all the pieces 
corresponding to the same composer into a single dataset.
%
%
Each dataset turns out to be constituted by a succession
of discretized chromas. 
Each repetition of a particular discretized chroma is a token of the corresponding codeword type.
The number of different types in the dataset 
(the types with absolute frequency greater than zero)
is then the vocabulary size, denoted by $V$
(this number will be smaller than 4096, in practice).
The sum of all the absolute frequencies of all types yields
the total number of tokens, which corresponds, by construction, 
to the dataset length $L$ measured in number of the selected time units.
In a formula, $\sum_{r=1}^V n_r =L$, 
where $r$ labels all the codeword types in the dataset and 
$n_r$ denotes the absolute frequency of type $r$.




\subsection{Musical corpus}

We perform our study over the {\it Kunstderfuge} corpus
\cite{Kunst}, 
which, at the time of our analysis, consisted of 17,418 MIDI files 
corresponding to pieces of 79 classical composers, from the 12th to the 20th century
(Ref. \cite{Lacasa_music} has scrutinized the {\it Kunstderfuge} corpus
looking for properties different than the ones we are interested in).
Removing files that were not clearly labelled
or corrupted (files that we are not able to process) 
and files for which we could not obtain the bar 
(and thefore could not determine the beat),
yields a remainder of 9489 files and 76 composers, 
ranging from Guillaume Dufay (1397-1474)
to
Messiaen (1908-1992).
The names of all composers are provided in Table \ref{tableonethree} (Appendix I).
The total length of the resulting corpus is 
$L=5,131,159$ tokens,
with a total vocabulary $V=4085$.

\begin{figure}[ht]
(a)
\includegraphics[width=.75\columnwidth]{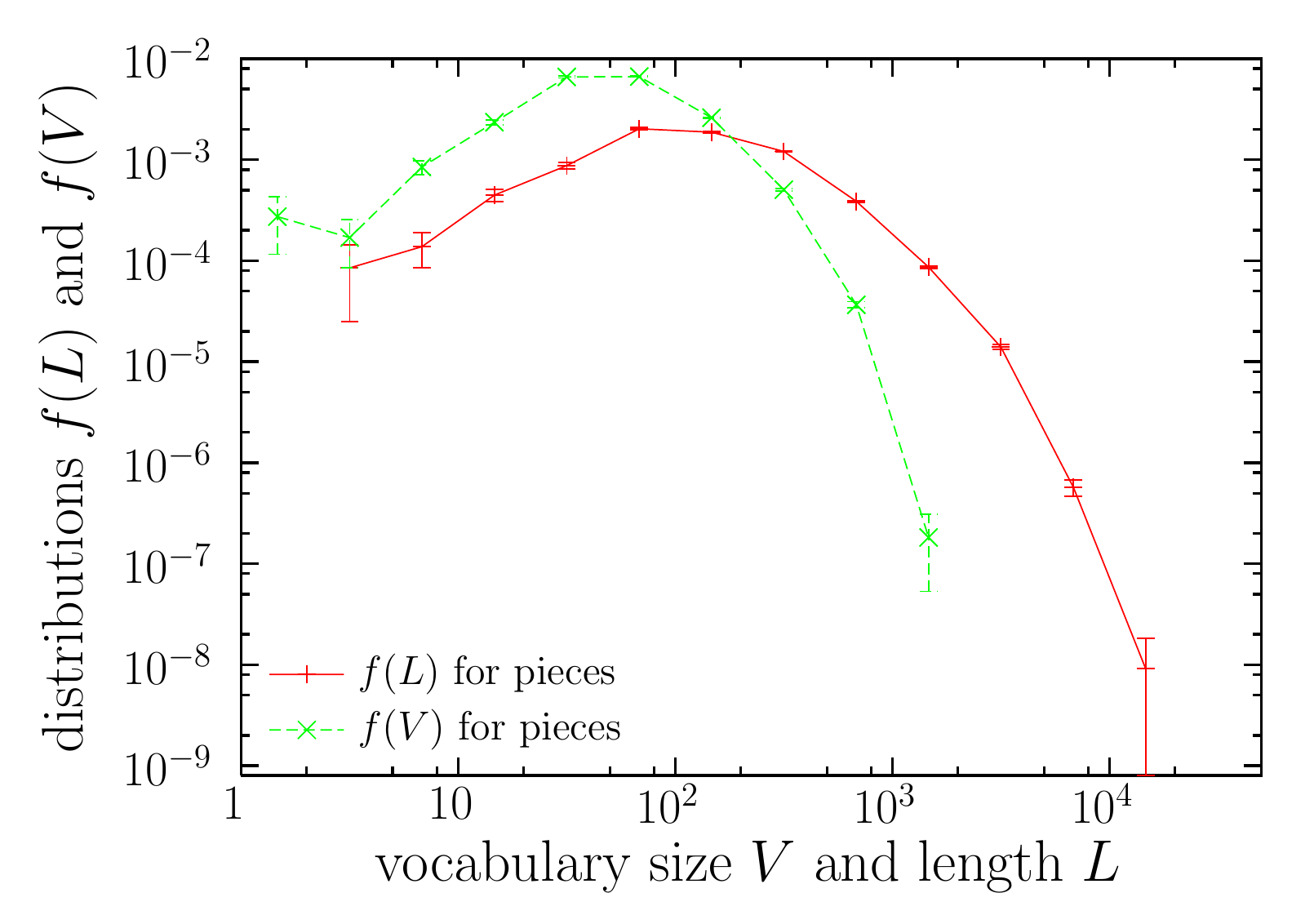}

(b)
\includegraphics[width=.75\columnwidth]{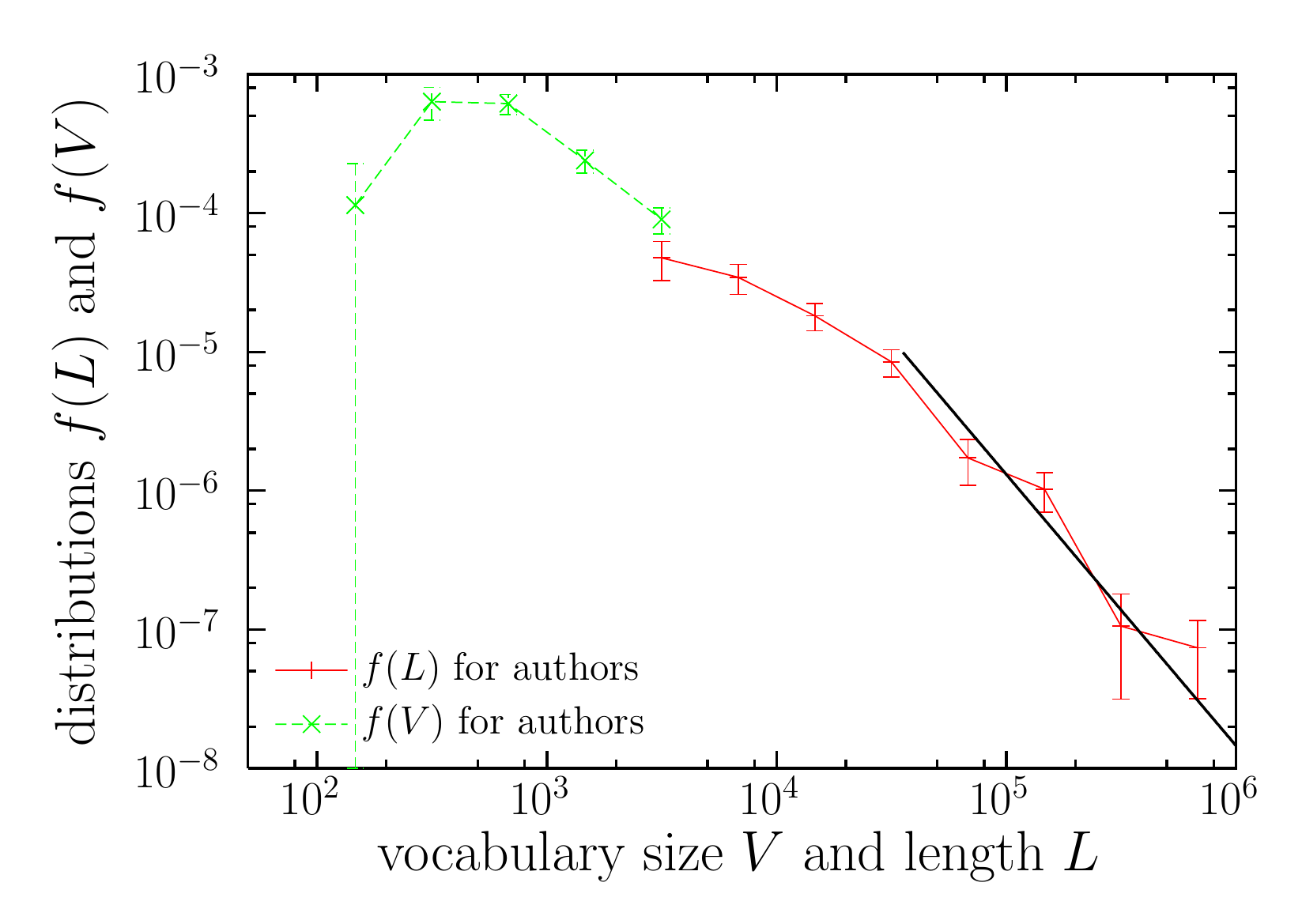}
\caption{
Probability mass functions of length $L$ and vocabulary size $V$. 
(a) For individual pieces.
(b) For individual authors. 
A power-law fit to the tail of $f(L)$ is shown as an indication (straight line), 
with an exponent $1.95\pm0.18$ 
(the fitting method is the one in Refs. \cite{Corral_Deluca,Corral_Gonzalez}).
}
\label{Fig_L_V_distr}
\end{figure}

Figure \ref{Fig_L_V_distr}(a) shows, for each piece, 
its length $L$ and the size of its vocabulary $V$, 
by means of the corresponding distributions (probability mass functions
$f(L)$ and $f(V)$). 
Both $L$ and $V$ show a variability of around three orders of magnitude,
with a maximum (i.e., a mode) around 100 tokens for $L$
and 50 types for $V$.
The distributions of $L$ and $V$ 
would correspond to 
intrinsic properties of classical musical compositions, 
which can nevertheless be biased due to subjective or arbitrary criteria employed when creating the {\it Kunstderfuge} corpus.


As mentioned, 
for statistical purposes
we aggregate all the compositions by the same author into a single dataset for that author.
Figure \ref{Fig_L_V_distr}(b) does the same job for authors as 
Figure \ref{Fig_L_V_distr}(a)
was doing for single pieces.
Since $L$ is a purely additive quantity (by definition of token), 
the values of $L$ increase when going from pieces to authors, 
but the variability of three orders of magnitude is more or less maintained.
The tail of $f(L)$ can be fitted by a decreasing power law, 
with an exponent around $1.95$, although the number of data points (composers)
in the tail is small. 
(in Ref. \cite{Moreno_Sanchez},
a power-law tail with an exponent close 3
was proposed for the distribution of text lengths,
but for individual written works, i.e., without author aggregation).  
In contrast to $L$, the size of vocabulary $V$ decreases its variability 
when going from pieces to composers, 
as types are not additive.
The distributions $f(L)$ and $f(V)$ at the author level do not show any 
musical characteristic, as we expect them to be incomplete, in principle 
(not all the pieces from each author are present in {\it Kunstderfuge}).
So, they characterize the corpus, and not the creativity of the composers.


%
%


\section{Fulfilment of Heaps' law} 

\begin{figure}[ht]
\includegraphics[width=.95\columnwidth]{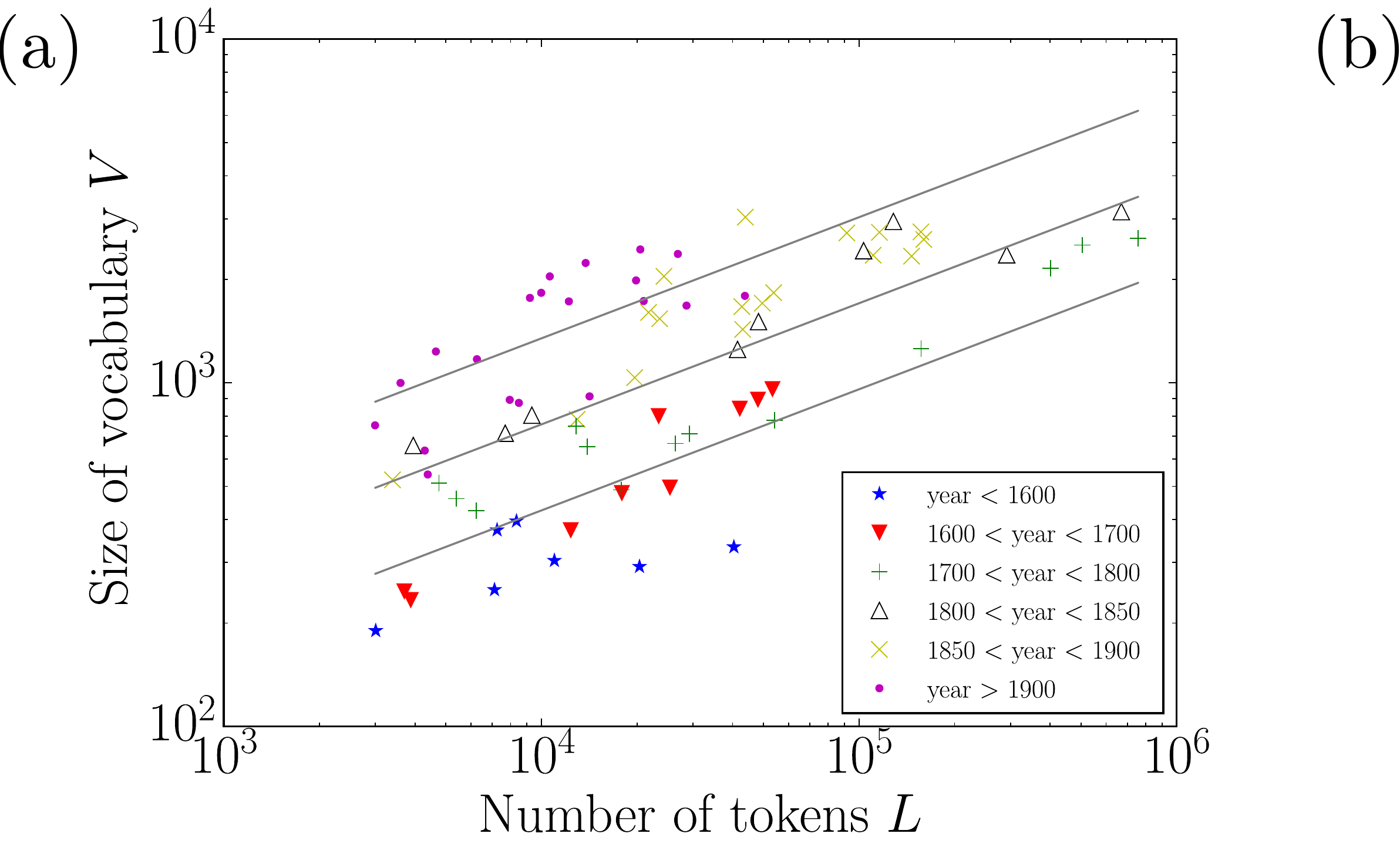}
\includegraphics[width=.8\columnwidth]{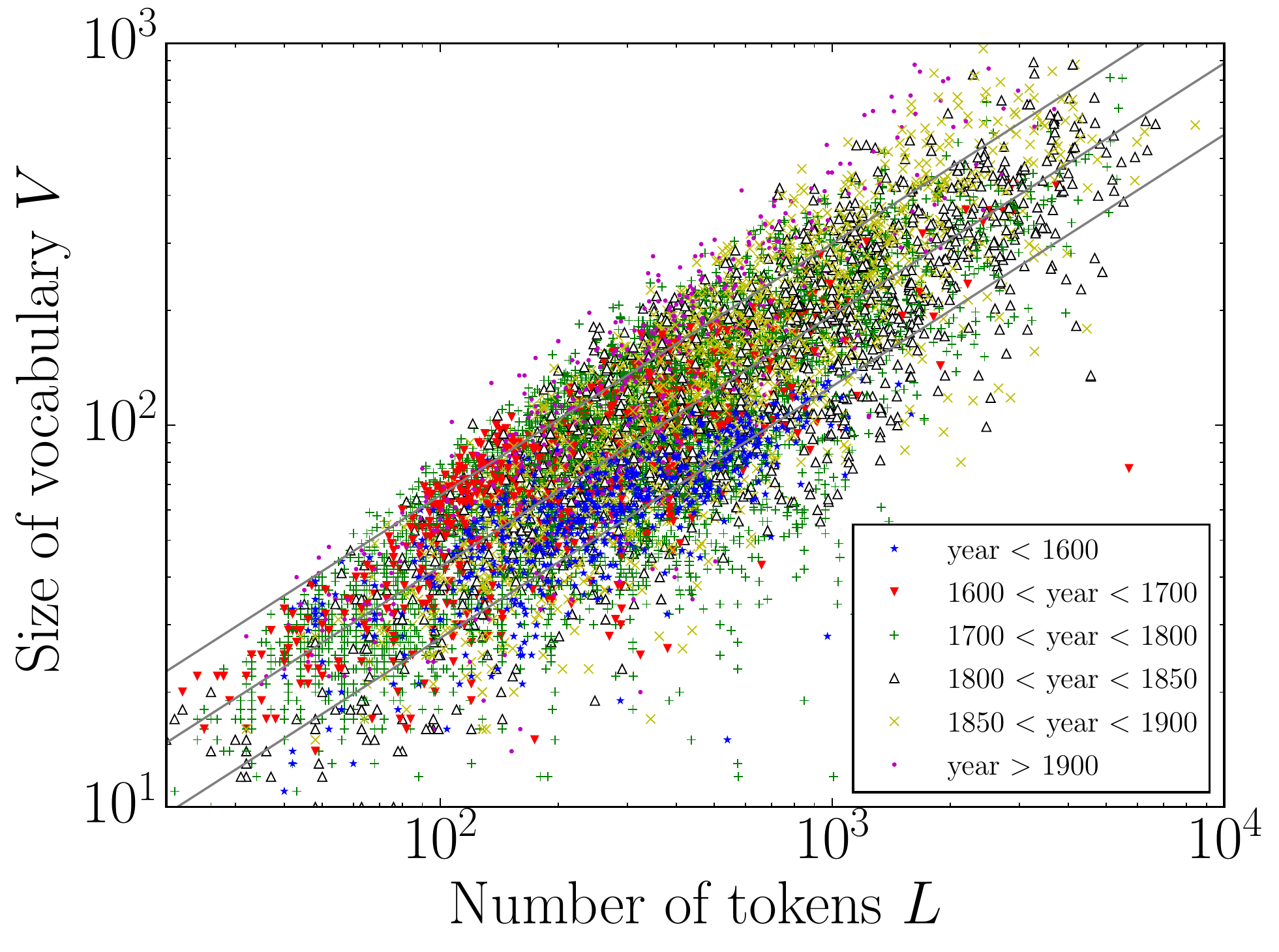}
\caption{
(a)
Scatter plot of size of vocabulary versus number of tokens (length)
for the 76 composers in the corpus, who are grouped chronologically, 
as represented by the points of different color
(one point is one composer).
The year of each composer is the mean between the birth year plus 20 and the death year.
Heaps' law is given by the central straight line
and the parallel lines denote one standard deviation $\sigma_c$.
Notice that when a limited chronological span is considered, 
the scattering is reduced.
(b) Analog scatter plot for the 9489 individual pieces.
Year of a piece is approximated to the year of its composer. 
The results of the fits are in Table \ref{tableonehalf} (in Appendix I).
}
\label{Fig_heaps}
\end{figure}

As an illustration of the data we deal with, 
we report in Table \ref{tableone} (in Appendix I)
the values of length $L$ and
size of vocabulary $V$
for the 15 composers 
with the largest values of $V$,
as derived from the {\it Kunstderfuge} corpus. 
All of them turn out to be very well-known composers, 
except perhaps 
Ferruccio Busoni (an Italian virtuoso pianist).
The top-3 most represented composers in terms of length $L$ are
Bach, Beethoven, and Mozart, 
with $L\simeq 760$,
$670$, and
$500$ thousand tokens, respectively. 
%
%
Figure \ref{Fig_heaps}(a) shows the corresponding scatter plot
for all 76 composers. 
The average increase of $V$ with $L$ is apparent, but with considerable scattering.
%


In order to obtain the parameters of Heaps' law,
we fit a regression line to the relation between $\log V$ and $\log L$, 
from which
we obtain a Heaps' exponent $\alpha\simeq 0.35$,
with a linear (Pearson) correlation coefficient 
$\rho\simeq 0.64$
(the value of $\log_{10}K$ turns out to be 1.47).
All the results of the fit are available in Table \ref{tableonehalf} (Appendix I),
and the resulting power law is represented in Fig. \ref{Fig_heaps}.
For completeness, we perform the same power-law fit for individual pieces,
see Fig. \ref{Fig_heaps}(b).
The results, also in Table \ref{tableonehalf}, 
show a much higher exponent, 
$\alpha \simeq 0.66$ 
(clearly above 0.5 this time), and 
also a higher linear correlation coefficient, $\rho\simeq 0.85$.

Thus, the supposedly universal values of $\alpha$ and $K$ 
for texts
mentioned in the introduction 
(either $\alpha=0.5$ or $K=1$ \cite{Wimmer_Altmann})
clearly do not hold for music, at least at the level of composers.
In any case, it is apparent that the larger the value of $L$, the larger (on average) the vocabulary size, but the increase in $V$ is rather modest, due to the small value of the exponent $\alpha$
(in other words, 
we need a 7 times longer piece for seeing a doubling of the vocabulary size).
The relative large value of the constant $K$ reported above for the composers
arises as a compensation for the small value of $\alpha$.

%



\section{Richness of vocabulary}



Due to the existence of Heaps' relation between $V$ and $L$,
as explained in the introduction,
it is erroneous to identify vocabulary size with vocabulary richness.
For each composer, 
Heaps' law
provides a natural way to correct the changes in $V$ 
due to the heterogeneities and biases in $L$.
We have also mentioned how the Giraud's and Herdan's indices 
are based on supposedly universal properties of Heaps' law.
However, notice that in the case of music, such universal properties do not hold anymore,
as $\alpha\ne 0.5$ and $K\ne 1$. 

Therefore,
we develop here a measurement of richness based on the empirical
(non-universal) validity of Heaps' law,
which will be relative to the rest of authors
in a given corpus.
In short, composers (or datasets, in general) 
with $V$ above the ``regression line'' in the scatter plot
(for the corresponding value of $L$) will have 
a vocabulary richness greater than what the Heaps' power law predicts, 
whereas composers below it will have lower richness.

Thus, we will calculate the difference 
$\log_{10} V - \log_{10} K -\alpha \log_{10} L$ 
from the empirical data
(using the fitted values of $K$ and $\alpha$),
and will rescale the result in ``units'' of $\sigma_c = \sigma_y \sqrt{1-\rho^2}$,
as the theory of linear regression tells us 
that the standard deviation of $\log_{10} V$ at fixed $L$
is $\sigma_c$, with $\sigma_y$ the standard deviation of $\log_{10} V$ for all values of $L$.
So, the vocabulary relative richness $R$ of each composer is defined as 
\begin{equation}
R=\log_{10} \left(\frac V {K L^\alpha}\right)^{1/\sigma_c}.
\label{equationforR}
\end{equation}
With that, 
$R>0$ will correspond to high vocabulary richness (higher than average)
whereas $R<0$ is, obviously, negative richness, i.e., 
poverty of vocabulary.
This easiness of interpretation of our
relative richness is an additional advantage in comparison to the indices of richness
$I_G$ and $I_H$, whose values are not directly interpretable.
The rescaling of the logarithm in terms of $\sigma_c$ 
in Eq. (\ref{equationforR}) is not necessary
if we deal with a single corpus (as it is the case here),
but could be useful to compare different corpora or even different 
phenomena, such as musical richness with literary richness.
Rescaling also provides further interpretability for the values of $R$,
in terms of standard deviations from the mean (in logarithmic scale).



When applied to our data, 
the composer with highest value of $R$ turns out to be Paul Hindemith 
($R=1.68$), 
closely followed by  Sergu\'ei Rajm\'aninov
(with nearly the same $R$). 
The lowest $R$ corresponds to 
Tom\'as Luis de Victoria 
($R=-2.49$).
In Table \ref{tablethree} (Appendix I) 
we provide some more details 
on the ranking of composers by different metrics.
General information on these composers can be found in 
Table \ref{tableonetwo} (Appendix I).
%
%

\begin{figure}[ht]
\includegraphics[width=.99\columnwidth]{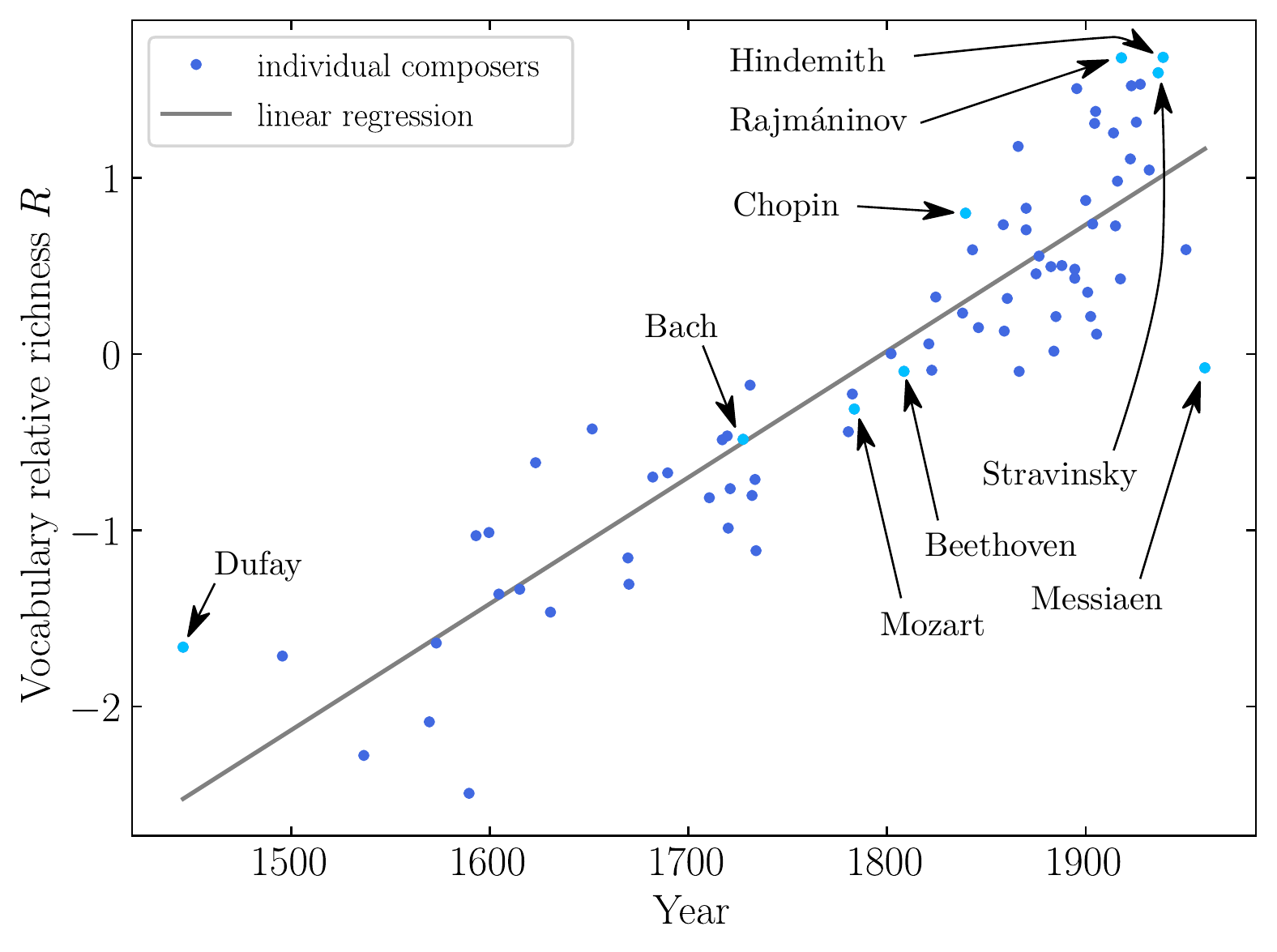}
\caption{
Vocabulary relative richness $R$ for each composer
in chronological order. 
Horizontal axis is birth $+$ $20$ $+$ death, divided by $2$. 
The straight line is a linear regression with slope $0.72\pm 0.04$ ``units of richness'' per century
and a linear correlation coefficient $\rho=0.90$
(and intercept $-12.9$). 
Some particular composers are highlighted, for the sake of illustration.
%
}
\label{Fig_trend}
\end{figure}

\begin{figure}[ht]
\includegraphics[width=.95\columnwidth]{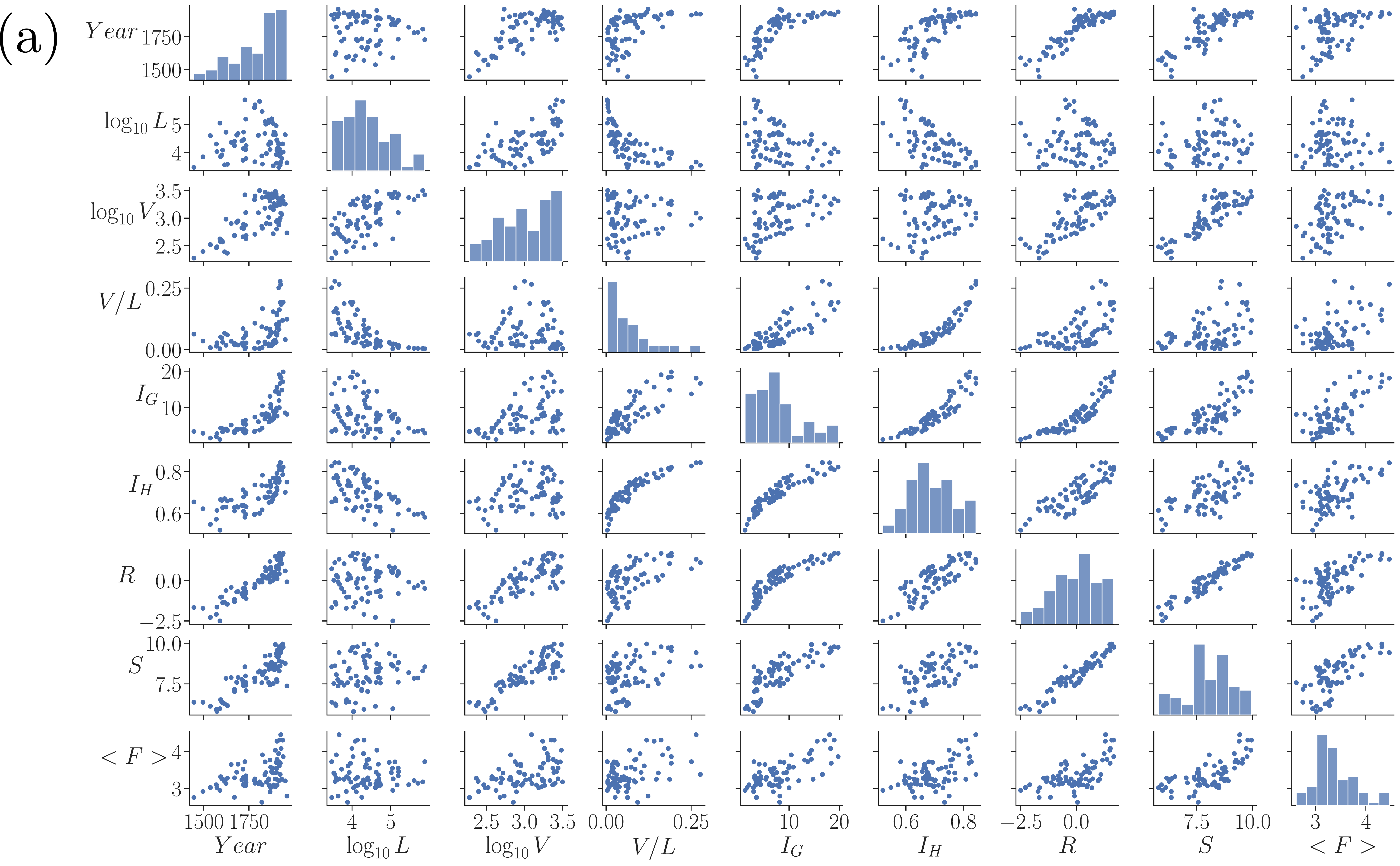}\\
\includegraphics[width=.95\columnwidth]{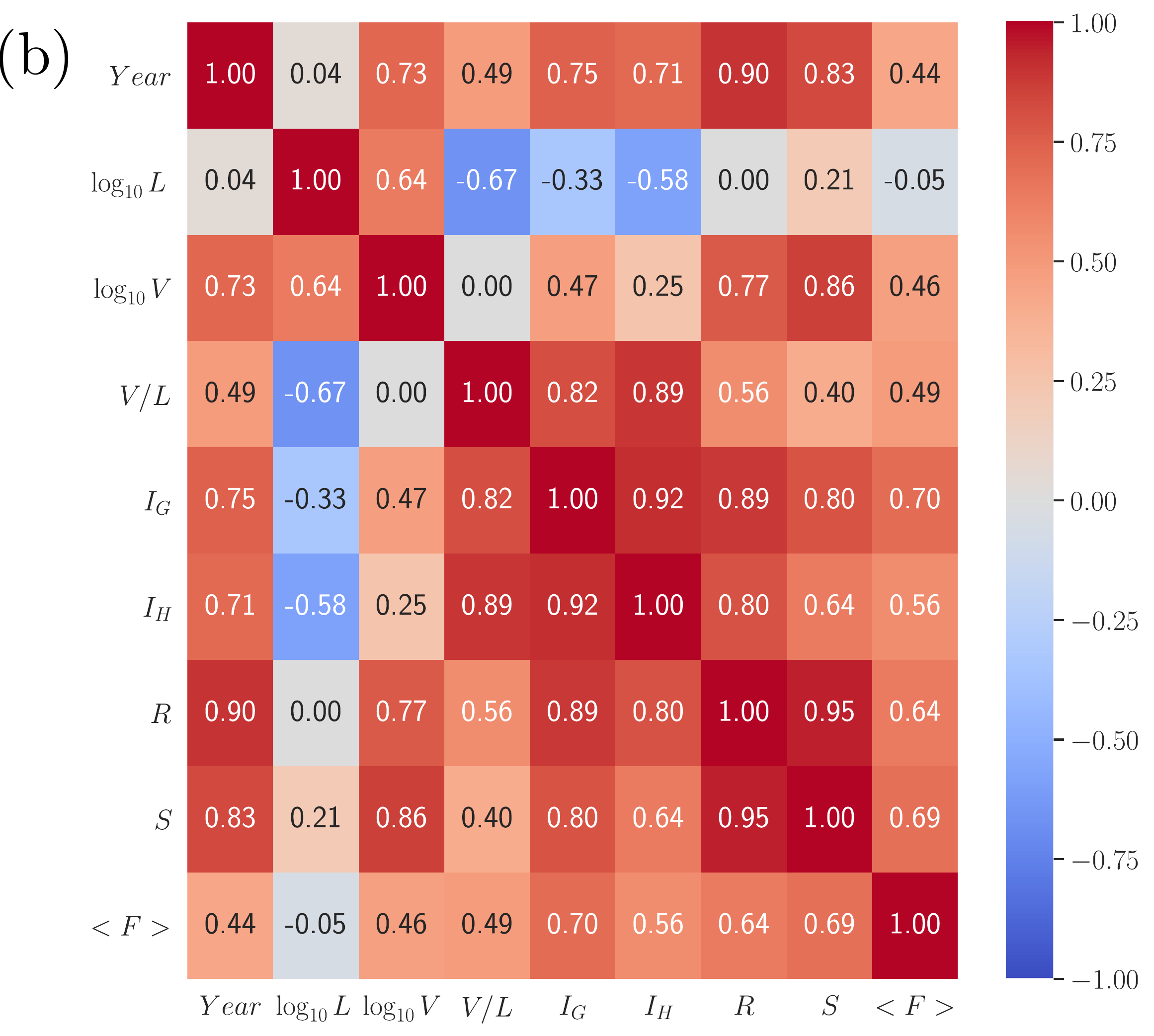}
\caption{
Comparison of the different metrics characterizing each composer.
(a) Scatter plots.
(b) Matrix of linear correlation coefficients. 
}
\label{Fig_scatter}
\end{figure}

Interestingly,
the chronological display 
in Fig. \ref{Fig_trend}
of the relative richness $R$ of  
each composer 
shows a clear increasing trend of this richness across the centuries, 
with a linear-slope increase of 0.72 ``units of richness'' per century
(and linear correlation coefficient $\rho=0.90$).
%
It is interesting to note that Ref. \cite{Serra_scirep} 
reported a decreasing trend for contemporary Western popular music; 
nevertheless, it should be also noted that, besides the different genres and epochs, 
richness was measured by modeling transitions between codewords using a complex network approach.
%
%
%
%
Figure \ref{Fig_scatter} compares our relative richness $R$, for each composer, 
with the logarithms of other composers' characteristics ($L$, $V$, type-token ratio, $I_G$, etc.).
Figure \ref{Fig_scatter}(a) shows all pairs of scatter plots between the metrics of the composers 
and
Fig. \ref{Fig_scatter}(b) shows the corresponding matrix of linear correlation coefficients.
The high correlation between the relative richness and the year of each composer is apparent (below we will discuss this figure in more detail).

%

\section{Comparison with entropy and filling of codewords}

A quantity of relevance in this context is the entropy of the distribution of 
type frequency~\cite{Corral_Cancho}. 
Each codeword type $r$ (with $r=1,2\dots V$) has a relative frequency 
given by its number of tokens divided by $L$, 
which we can assimilate to the probability of the type $P_r$
in the selected dataset, which in our case corresponds to individual composers.
The Shannon entropy (in bits) characterizing the vocabulary of each composer is simply 
$$S=-\sum_{r=1}^V P_r \log_2 P_r.$$
In the hypothetical (unrealistic) case of uniform use of all possible types, 
$V$ would reach its maximum value, $2^{12}$,
and we would obtain the maximum possible entropy, $S_{max}=\log_2 2^{12}=12$ bits.

In our composers' corpus, 
the highest entropy is given by
Stravinsky ($S=9.95$ bits),
followed by Gustav Mahler 
($S=9.91$ bits).
The lowest value of entropy corresponds
to Orlande de Lassus 
(or Orlando di Lasso, 
with $S$ around 5.8 bits).
%
%
We see that, as in the relative richness, 
the composers with the lowest entropies are from the 15th or 16th centuries, 
whereas those with the highest entropies are from the end of the 19th or the 20th century. 
The systematic increase of the entropy when the composers 
are ordered chronologically is analogous to the increase shown in Fig. \ref{Fig_trend}.
Linear regression shows that the increase in entropy is 0.69 bits per century
(with a linear correlation coefficient equal to 0.83).
Thus, the history of classical music (on average)
seems to extend the law of the increase of entropy with time beyond the 
usual physical systems considered in thermodynamics
\cite{Ben_Naim}.
The comparison of the entropy with the other metrics is included in Fig. \ref{Fig_scatter}.

Finally,
another metric of potential interest is the average filling of the codeword types
for each composer, 
calculated as $\langle F \rangle = \sum_{r=1}^V P_r F_r$, with $F_r$ the number of ones
in the discretized chroma corresponding to codeword type $r$
(if the codeword $r$ is represented as a 12-component vector, then $F_r=\sum_{i=0}^{11} r_i$, with $r_i$ the $i$-th component of the vector $r$, taking values $r_i=0$ or 1, as we have defined).
The composer with more average filling is 
Nikolai Medtner ($\langle F\rangle=4.46$), 
and the one with less value 
is Niccol\`o Paganini
(the most ancient composer in the corpus, with $\langle F\rangle=2.62$).
This metric also shows an increasing trend across the years, 
but, in contrast to $R$ and $S$, its correlation with time is not high.
The scatter plot, included in Fig. \ref{Fig_scatter}(a), 
has an upper envelope that increases linearly in time,
but the lower envelope is rather constant.
In this way, Messiaen (the most modern composer included in the corpus),
shows one of the lowest values of $\langle F \rangle$.
The average filling is compared with the rest of metrics also in Fig. \ref{Fig_scatter}(b),
where we see that it is not very highly correlated with any of them.

\section{Robustness of results}

We have tested the robustness of our results in front of the arbitrariness contained in the process of codeword construction.
For that purpose, we have investigated the effect of changing the value of the discretization threshold.
We find that thresholds ranging from 0.025 to 0.2
lead to very minor changes of the Heaps' exponent $\alpha$,
with values comprised between 0.350 and 0.355.
A threshold equal to zero leads to $\alpha$ around 0.355.
Thresholds higher than 0.3 lead to progressively increasing $\alpha$, 
e.g., for 0.5 we get $\alpha\simeq 0.39$.
The intercept in Heaps' regression ($K$) shows higher variability;
nevertheless, although the specific values of the relative richness $R$ 
depend on $K$, their statistical properties do not change
(as $K$ is only bringing a shift in logarithmic scale).
The stability of $\alpha$ ensures the robustness of $R$.

We have also explored the effects of changing the time unit over which the
codewords are built,
finding that an increase in the time unit from 0.5 to 1.5 beats leads to 
a modest increase in $\alpha$ from 0.33 to 0.39 (with very little influence in the value of the threshold, 
in the same way as explained above).
Taking a time unit as large as 4 beats leads to $\alpha$ close to 0.5.
This is a considerable change, but not unexpected, 
as, for instance, in a $^4_4$ bar, 4 beats correspond to one bar
and conventional musical knowledge tells us that 
bar-based codewords are entities with different properties than
beat-based codewords.

\section{Summary and discussion}

Summarizing the results,
the highest linear correlation between all the metrics characterizing the composers
is the one relating
the relative richness $R$ with the entropy 
(around 0.95, Fig. \ref{Fig_scatter}(b)),
but $R$ is also highly correlated with the Giraud index.
The Herdan index is also highly correlated with Giraud and with the type-token ratio.
The correlations obtained between $\langle F\rangle$ and the other indices are not so high.
Interestingly, the highest correlation of the year characterizing each composer is with the proposed relative richness $R$ 
(taking a value of 0.90, Fig. \ref{Fig_scatter}(b)).
The correlation of $R$ with $\log L$ is zero by construction.
Replacing Pearson linear correlation with Spearman or Kendall correlations 
does not qualitatively change very much this pattern of correlations (not shown).

So, we conclude that relative richness presents very interesting properties,
as it shows maximum correlation with entropy and with year.
Nevertheless, the correlation of relative richness with entropy is so high that one could consider instead
to use entropy to account for vocabulary richness.
This has the advantage that one does not need to have a whole corpus to obtain the Heaps' law in advance; rather, it can be calculated even for a single author or piece,
or it can be used for the comparison of just two authors.
Nonetheless, the entropy has the disadvantage that
one does not know what a high entropy or a low entropy is, in principle,
and thus, the apparent advantage of entropy is lost in a practical situation. 
Interpretability is further hampered 
by not having a direct notion of ``above$/$below average'' richness, 
which is corpus dependent.
From the comparison with the relative richness we see that the value of $S$ separating low richness and high richness is around 8 bits 
(for the present codification of 12 bits per codeword and for the present corpus). 
Very low richness is around 6 bits and very high around 10.
This also corresponds to an average filling of codewords (mean number of ones)
ranging from 
2.7 to 4.5. 
Another disadvantage of entropy is that one needs to count the repetitions of all codeword types (the $V$ values of $P_r$), 
whereas for the richness one only needs to know how many types there are
(the value of $V$).


In conclusion, 
we have analyzed 
the harmonic content of a large corpus of classical music in MIDI form.
The definition of music codewords allows us to quantify the size of the harmonic vocabulary of classical composers, and relate it to their musical productivity
(the total length of their compositions, 
as contained in the considered corpus).
We obtain that the relation between the two variables is well described 
by an increasing power law, which is analogous to the Heaps' law 
previously found in linguistics and other Zipfian-like systems.
Nevertheless, the obtained power-law exponent turns out to be somewhat small
($\alpha \simeq 0.35$), 
if we compare it to typical values found in linguistics.

Heaps' law allows us to develop a proper measure of vocabulary richness for each composer in relation to the rest of the corpus.
Remarkably, 
we find that vocabulary richness undergoes a clear increasing trend across 
the history of music, as expected from qualitative musical wisdom.
Our approach provides a transparent quantification of this phenomenon.
We also show that our relative richness is highly correlated with the entropy
of the distribution of codeword frequencies, 
so entropy can be equally used to measure vocabulary richness,
once it is properly calibrated.
%
Our metric has several advantages with respect to previous indices of richness, 
such as being relative to the richness of the rest of composers and a better interpretability of the values.
At our level of resolution (that of individual composers) 
the evolution of richness does not show
any revolutions or sudden jumps; 
instead, it seems to be rather gradual, 
and well fitted by a linear increase. 
(the variability is too high to allow one to obtain meaningful results
beyond the linear increasing trend).




\section{Appendix I: Tables}


In this appendix we include the tables cited in the text.
Table \ref{tableonethree} contains the names of the 76 composers in the corpus, in chronological order.
Table \ref{tableone} provides the values of $L$ and $V$ (together with other relevant figures)
for the 15 composers best represented (in terms of $V$) in the corpus.
Table \ref{tableonehalf} yields the results of fitting Heaps' law to the values of $V$ and $L$ for the individual composers and for the individual pieces.
The results for the composers without performing any transposition
are also included, for the sake of comparison.
Observe how the exponent $\alpha$ remains essentially the same.
Table \ref{tablethree} shows the top-5, medium-6, and bottom-5 composers regarding relative richness, entropy, and average filling.
Table \ref{tableonetwo} gives the general details of the composers highlighted in the previous table.

\begin{table}[H]
\begin{center}
\caption{\label{tableonethree}
Name of the 76 classical composers in the 
{\it Kunstderfuge} corpus analyzed in this paper.
The order is chronological (established by the year of birth).
}
\smallskip
\tiny
\begin{tabular}{|llll|}
\hline
G. Dufay & 
J. Desprez & 
C. de Morales & 
G. P. da Palestrina \\ 
O. Lassus & 
W. Byrd & 
T. L. de Victoria & 
J. Dowland \\
C. Gesualdo & 
C. Monteverdi & 
G. Frescobaldi & 
S. Scheidt \\
J. J. Froberger & 
J.-H. d'Anglebert &
J. B. Lully & 
D. Buxtehude \\ 
J. Pachelbel & 
F. Couperin & 
T. Albinoni & 
A. Vivaldi \\
G. P. Telemann & 
J.-F. Dandrieu & 
J.-P. Rameau & 
J. S. Bach \\ 
G. F. H\"andel & 
D. Scarlatti & 
D. Zipoli & 
J. Haydn \\
J. G. Albrechtsberger & 
M. Clementi & 
W. A. Mozart & 
L. van Beethoven \\ 
J. B. Cramer & 
N. Paganini & 
F. Schubert & 
H. Berlioz \\
F. Mendelssohn & 
F. Chopin & 
R. Schumann & 
F. Liszt \\ 
C.-V. Alkan & 
C. Franck & 
A. Bruckner & 
L. M. Gottschalk \\ 
J. Brahms & 
C. Saint-Sa\"ens & 
A. Guilmant & 
G. Bizet \\ 
M. M\'ussorgsky & 
P. I. Tchaikovsky & 
A. Dvo\v{r}\'ak & 
E. Grieg \\ 
G. U. Faur\'e & 
L. Jan\'a\v{c}ek & 
I. Alb\'eniz & 
G. Mahler \\ 
C. Debussy & 
F. Busoni & 
E. Satie & 
S. Joplin \\ 
L. Godowsky & 
A. Scriabin & 
S. Rajm\'aninov & 
M. Reger \\ 
A. Schoenberg & 
M. Ravel & 
S. Karg-Elert & 
O. Respighi \\ 
N. M\'edtner & 
B. Bart\'ok & 
\'I. Stravinsky & 
S. Prok\'ofiev \\ 
P. Hindemith & 
G. Gershwin & 
D. Shostak\'ovich & 
O. Messiaen\\
\hline
\end{tabular}
\par
\end{center}
\end{table}

\begin{table}[H]
\begin{center}
\caption{\label{tableone}
Length $L$ in number of tokens
and size of vocabulary $V$ for the 15 composers 
that are best represented in the {\it Kunstderfuge} corpus
(after transposition).
Year of birth, year of death, 
and number of pieces (\# pieces) are also included.
%
}
\smallskip
\footnotesize
\begin{tabular}{ |rlrrrrr|}
\hline
&Composer & Birth & Death & $L$ \, & $V$ \, & \# pieces\\ 
\hline
 & Johann Sebastian Bach & 1685  & 1750 &
757,945 &	2630 & 	2169\\
 & Joseph Haydn & 1732 & 1809 & 
401,162	& 2155 &	641\\
 & Wolfgang A. Mozart & 1756  & 1791 &  
504,386	& 2516 &	685\\
 & Ludwig van Beethoven & 1770  & 1827 &  
670,380 & 3146 &	638\\
 & Franz Schubert & 1797  & 1828 &
292,007 & 2356 & 270\\
 & Fr\'ed\'eric Chopin & 1810 & 1849  & 
 128,373	& 2947 &	220\\
 & Robert Schumann & 1810 & 1856 & 
103,432 & 2423 &	109\\
& Franz Liszt & 1811  & 1886 &
 116,044 & 2739 & 136\\
& Johannes Brahms & 1833 & 1897 &  
159,998 & 2612 &	148\\
& Piotr I. Tchaikovsky & 1840 & 1893 & 
 156,749	& 2748 &252\\
& Anton\'{\i}n Dvo\v{r}\'ak & 1841 & 1904 &
  110,855	& 2350 &145\\
& Gustav Mahler & 1860 & 1911 & 
 43,877 & 3034 &	33\\
& Claude Debussy & 1862  & 1918 &
 91,600 & 2728 &	194\\
& Ferruccio Busoni &	1866 &	1924 &	26,898 &	2370 &	42\\
& \'Igor Stravinsky & 1882 & 1971 & 
20,481 &	2443 &	39\\
\hline
& All 76 & 1397 & 1992 & 5,131,159  & 4085 & 9489        \\
\hline
\end{tabular}
\par
\end{center}
\end{table}

\begin{table}[H]
\begin{center}
\caption{\label{tableonehalf}
Results of fitting a regression line to the scatter plot between $\log_{10} V$ and $\log_{10}L$, 
as $\log_{10} V =\log_{10} K + \alpha \log_{10}L$ (Heaps' law),
with correlation coefficient $\rho$.
The value of $\sigma_c$ is also shown. 
Three cases are compared: 
composers with transposition (our main case of study),
composers without transposition (for the sake of comparison),
and individual pieces (for which transposition has no effect on $L$ and $V$).
%
%
}
\smallskip
\begin{tabular}{|l | cccc|}
\hline
Dataset &$\alpha$&$\log_{10} K$ & $\rho$ & $\sigma_c$\\
\hline
authors, transposed  & 
0.35 $\pm$  0.05 &
1.47 &
0.64 &
0.25 \\
authors, no transposed  &
0.35 $\pm$  0.05 &
1.53&
0.67&
0.23 \\
pieces   &
0.659 $\pm$  0.004 &
0.31 & 
0.85 & 
0.19 \\
\hline
\end{tabular}
\par
\end{center}
\end{table}


\begin{table}[H]
\begin{center}
\caption{\label{tablethree}
Top-5, middle-6, and bottom-5 composers
as ranked by relative richness $R$, 
entropy $S$,
and average filling $\langle F \rangle$.
}
\smallskip
\footnotesize
\begin{tabular}{|lr|lr|lr|}
\hline
Rank by $R$ & $R$ & Rank by $S$ & $S$ (bits) & Rank by $\langle F\rangle $ & $\langle F \rangle$ \\

\hline
Hindemith	&	1.683	&	Stravinsky	&	9.945	&	Medtner &	4.462	\\
Rajm\'aninov	&	1.680	&	Mahler	&	9.914	&	Rajm\'aninov &	4.316 \\
Stravinsky	&	1.595	&	Rajm\'aninov	&	9.777	&	Stravinsky &	4.314 \\
Gershwin	&	1.530	&	Hindemith	&	9.746	&	Reger	& 4.285\\
Bart\'ok	&	1.521	&	Gershwin	&	9.709	&	Gershwin	& 4.087\\
		\hline									
Grieg	&	0.212	&	Mendelssohn	&	8.265	&	Scheidt &	3.277	\\
Berlioz	& 0.150 & Clementi &	8.222 & Gesualdo	& 3.266\\
Gottschalk &	0.130 & Schubert & 	8.190& Liszt	& 3.258\\
Satie	 & 0.112 & Cramer	& 8.065 & Albrechtsberger &	3.257\\
Paganini &	0.057 & Scarlatti	& 7.984  & Faur\'e	& 3.256\\
Guilmant &	0.016 & Pachelbel	& 7.898 & Dowland	& 3.255\\
				\hline							
Dufay	&	-1.663	&	Scheidt	&	6.169	&	Bizet	 & 2.906	\\
Desprez	&	-1.714	&	Morales	&	6.128	&	Messiaen	& 2.791	\\
Palestrina	&	-2.087	&	Victoria	&	5.982	&	Lully	& 2.756	\\
Morales	&	-2.277	&	Palestrina	&	5.938	&	Dufay &	2.751	\\
Victoria	&	-2.492	&	Lassus	&	5.803	&	Paganini &	2.621	\\
\hline
\end{tabular}
\par
\end{center}
\end{table}

\begin{table}[H]
\begin{center}
\caption{\label{tableonetwo}
As Table \ref{tableone}, 
but for the composers that appear in Table \ref{tablethree}
(and do not appear in the former), in chronological order. 
The horizontal line separates composers in the top-5, middle-6, and bottom-5 categories 
for relative richness and entropy (not for $\langle F \rangle$).
}
\smallskip
\scriptsize
\begin{tabular}{ |lrrrrr|}
\hline
Composer & Birth & Death & $L$ \, & $V$ \, & \# pieces\\ 
\hline
Guillaume Dufay	&	1397	&	1474	&	3008	&	190	&	10	\\
Josquin Desprez	&	1450	&	1521	&	7118	&	250	&	32	\\
Crist\'obal de Morales	&	1500	&	1553	&	40,380	&	333	&	88	\\
Giovanni P. da Palestrina	&	1525	&	1594	&	20,369	&	292	&	70	\\
Orlande de Lassus	&	1532	&	1594	&	10,988	&	304	&	70	\\
Tom\'as Luis de Victoria	&	1548	&	1611	&	113,162	&	423	&	333	\\
John Dowland	&	1563	&	1626	&	12,376	&	372	&	61	\\  
Carlo Gesualdo	&	1566	&	1613	&	8347	&	396	&	37	\\ 
Samuel Scheidt	&	1587	&	1654	&	3878	&	233	&	8	\\
Jean-Baptiste Lully	&	1632	&	1687	&	17,917	&	477	&	107	\\
\hline
Johann Pachelbel	&	1653	&	1706	&	48,081	&	892	&	182	\\
Domenico Scarlatti	&	1685	&	1757	&	12,874	&	747	&	43	\\
Johann Georg Albrechtsberger	&	1736	&	1809	&	4760	&	511	&	13	\\
Muzio Clementi	&	1752	&	1832	&	41,458	&	1250	&	46	\\
Johann Baptist Cramer	&	1771	&	1858	&	3949	&	657	&	29	\\
Niccol\`o Paganini	&	1782	&	1840	&	7692	&	713	&	23	\\						
Hector Berlioz	&	1803	&	1869	&	9333	&	805	&	8	\\
Felix Mendelssohn 	&	1809	&	1847	&	48,262	&	1506	&	50	\\
Louis Moreau Gottschalk	&	1829	&	1869	&	19,675	&	1035	&	36	\\						
Alexandre Guilmant	&	1837	&	1911	&	3397	&	522	&	11	\\
Georges Bizet	&	1838	&	1875	&	12,955	&	783	&	21	\\ 
Edvard Grieg	&	1843	&	1907	&	43,033	&	1430	&	30	\\
Gabriel Urbain Faur\'e	&	1845	&	1924	&	49,608	&	1704	&	97	\\
Erik Satie	&	1866	&	1925	&	14,179	&	913	&	47	\\						
\hline
Max Reger	&	1873	&	1916	&	12,205	&	1725	&	32	\\
Sergu\'ei Rajm\'aninov	&	1873	&	1943	&	13,783	&	2231	&	16	\\
Nikolai Medtner	&	1880	&	1951	&	4653	&	1233	&	7	\\
B\'ela Bart\'ok 	&	1881	&	1945	&	9203	&	1766	&	18	\\
Paul Hindemith	&	1895	&	1963	&	10,626	&	2039	&	22	\\
George Gershwin	&	1898	&	1937	&	9988	&	1827	&	12	\\
Olivier Messiaen	&	1908	&	1992	&	4386	&	541	&	9	\\
\hline
\end{tabular}
\par
\end{center}
\end{table}

\section{Appendix II: Transposition procedure}

The method we use for the analysis of the key 
is the Krumhansl-Schmuckler Key-Finding Algorithm 
\cite{temperley1999}. 
This algorithm is based on the key profiles described in Ref. \cite{Krumhansl1982},
which were obtained by empirical experiments where subjects rated how well each pitch fitted a prior context establishing a key.  The values for the major key profile are 
6.35, 2.23, 3.48, 2.33, 4.38, 4.09, 2.52, 5.19, 2.39, 3.66, 2.29 and 2.88, 
where the first number corresponds to the mean rating for the tonic of the key, the second to the next of the 12 tones in the chromatic scale, etc. 
The values for the minor key context are 
6.33, 2.68, 3.52, 5.38, 2.60, 3.53, 2.54, 4.75, 3.98, 2.69, 3.34 and 3.17 
\cite{Krumhansl2010bis}.

\begin{figure}[H]
\includegraphics[width=.95\columnwidth]{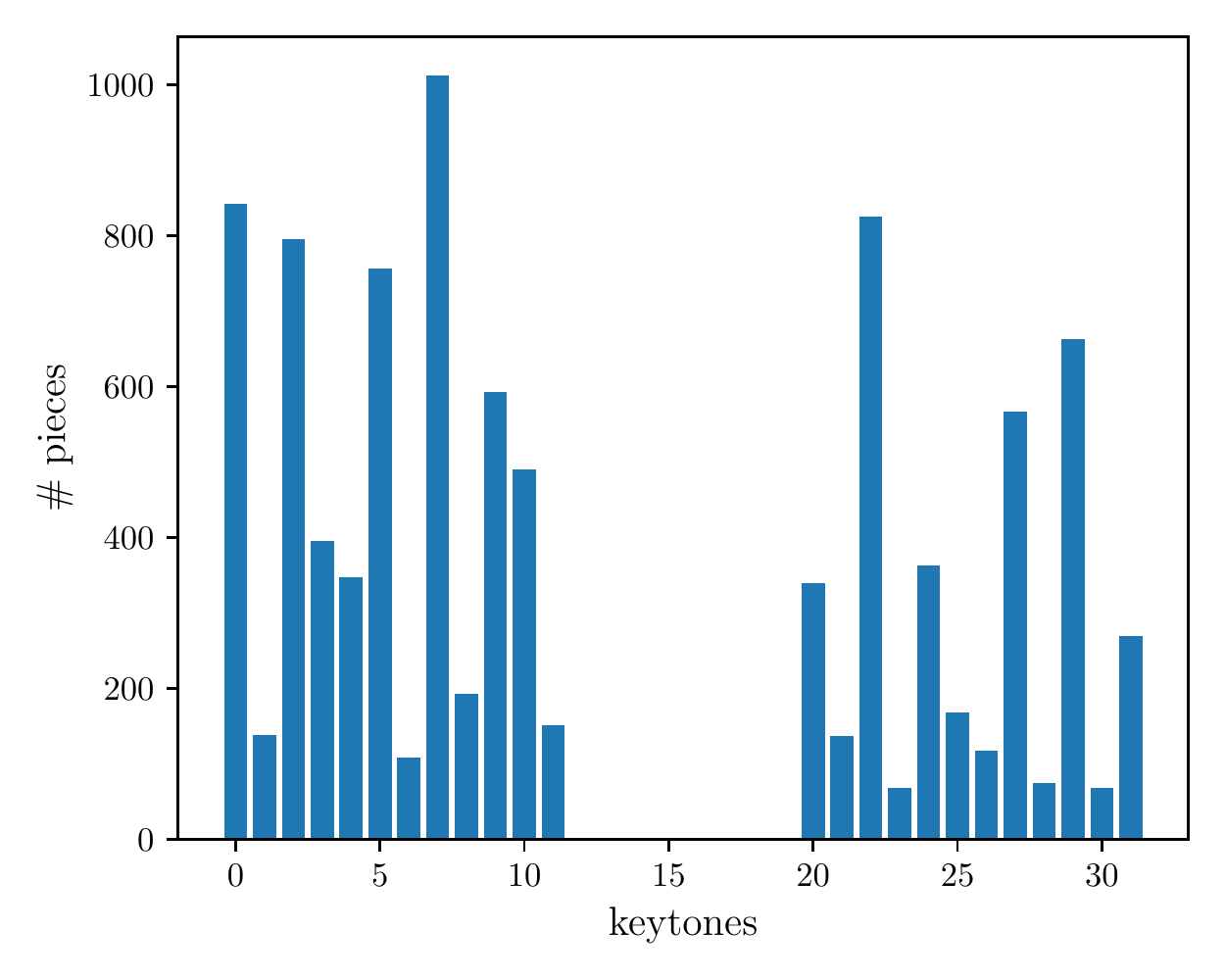}
\caption{
Absolute abundance of each key in the corpus, 
counted in number of pieces,
before transposition, obviously.
Zero corresponds to C Major, one to C\#/D$\flat$ Major... up to B Major;
20 corresponds to C minor and so on.
}
\label{Figkeys}
\end{figure}

The procedure of the algorithm to calculate the key of a piece 
is as follows:
\begin{enumerate}
\item Average, in terms of tokens, 
the discretized chromas in the considered piece.
\item 
Calculate the Pearson linear correlation of the major key profiles 
with the average discretized chroma and all their 
circular shifts (12 values).
\item Repeat step 2 using the minor key profiles.
\item The transposition shift is the shift that maximizes the correlation from both steps 2 and 3.
\end{enumerate}
The key is obtained directly from the transposition shift, with the following correspondence:
$0 = C$, $1 = C\#/D\flat$, $2 = D$, $3 = D\#/E\flat$ \dots
For the {\it Kunstderfuge} corpus we find that the most common keys are
G Major, C Major, and D minor,
as shown in Fig. \ref{Figkeys},  
whereas the least common are E$\flat$ minor and B$\flat$ minor.
%

We do not use the information about the key provided by the metadata 
in the MIDI file, as this seems to refer to the key signature rather than to the key.
To check that the transposition is being done correctly, we verify that the most common
codewords after transposition are $CEG$, $CFA$, $CEA$, $DGB$, and the empty codeword, 
corresponding to a silent beat. 
Table \ref{tablekeys} shows the top 10 codewords, in terms of absolute frequency, 
making clear how these can be related to chords or 
tonal functions that are common in C major or A minor.
in $C$ Major or $A$ minor.

\begin{table}[h]
\begin{center}
\caption{\label{tablekeys}
Ten most common codewords in the corpus (after transposition), 
in terms of number of tokens (absolute frequency).
}
\smallskip
\begin{tabular}{|lllr|}
\hline
Codeword & Chord && Frequency\\
\hline
100010010000 &CEG &I &257,252\\
100001000100 &CFA & IV&145,967\\
100010000100 &CEA &vi &119,734\\

001000010001 &DGB &V& 105,361\\
000000000000 &&&99,761\\
100010000000 &CE &I&86,179\\
100000000000 &C &I&78,009\\
001001000100 &DFA &ii&75,462\\
001001010001 &DFGB &V&70,802\\
000000010000 &G &V& 58,966\\
\hline
\end{tabular}
\par
\end{center}
\end{table}

%

\section{Acknowledgements}

Some preliminary work of this project was done by I. Moreno-S\'anchez,
funded by the Collaborative Mathematics Project of the La Caixa Foundation.
Also, 
support from projects
FIS2015-71851-P and
PGC-FIS2018-099629-B-I00
from Spanish MINECO and MICINN
is acknowledged.
M. S.-P.'s participation has been possible thanks to the Internship Program of
the Centre de Recerca Matem\`atica.


\section{Authors contributions} 

M. S.-P. wrote the code and analyzed the data,
J. S. and A. C. supervised the results,
all authors interpreted the results.
A. C. wrote the first draft of the manuscript and
all authors revised the manuscript.

\section{Competing interests}
The authors declare no competing interests.






%
%








%


\end{document}